\documentclass[preprint,noshowpacs,preprintnumbers,amsmath,amssymb]{revtex4}

\usepackage[utf8]{inputenc}
\usepackage{amsfonts,amsmath,amssymb}
\usepackage{graphicx}
\usepackage{bm} 
\usepackage{mathrsfs}
\usepackage{textcomp}
\usepackage{subfigure}
\usepackage[version=3]{mhchem}
\usepackage{hyperref}
\usepackage[per-mode = symbol]{siunitx}
\usepackage{xspace}
\let\orgautoref\autoref
\providecommand{\Autoref}{%
  \def\sectionautorefname{Section}%
  \def\figureautorefname{Figure}%
  \def\subfigureautorefname{Figure}%
  \orgautoref}
\renewcommand{\autoref}{%
  \def\sectionautorefname{Sec.}%
  \def\figureautorefname{Fig.}%
  \def\subfigureautorefname{Fig.}%
  \orgautoref}

\newcommand{\cost}{\ensuremath{\langle\cos^2\theta_\text{2D}\rangle}\xspace}
\newcommand{\cosa}{\ensuremath{\langle\cos^2\alpha\rangle}\xspace}

\newlength{\figwidth}
\newlength{\figwidthsmall}
\begin{document}


\title{Long-lasting field-free alignment of large molecules inside helium nanodroplets}

\author{Adam S. Chatterley}%
\author{Constant Schouder}%
\author{Lars Christiansen}%
\author{Benjamin Shepperson}%
\author{Mette Heidemann Rasmussen}%
\author{Henrik Stapelfeldt}%
\email[]{henriks@chem.au.dk}%
\affiliation{Department of Chemistry, Aarhus University, 8000 Aarhus C, Denmark}

\date{\today}

\maketitle



\textbf{Molecules with their axes sharply confined in space, available through laser-induced alignment methods~\cite{stapelfeldt_colloquium:_2003,fleischer_molecular_2012}, are essential for many current experiments, including ultrafast molecular imaging~\cite{hockett_time-resolved_2011,christensen_dynamic_2014,kraus_measurement_2015,wolter_ultrafast_2016}. Most of these applications require both that the aligning laser field is turned-off, to avoid undesired perturbations, and that the molecules remain aligned sufficiently long that reactions and dynamics can be mapped out. Presently, this is only possible for small, linear molecules and for times less than 1 picosecond. Here, we demonstrate strong, field-free alignment of large molecules inside helium nanodroplets, lasting tens of picoseconds. Molecular alignment in either one or three dimensions is created by a slowly switched-on laser pulse, made field-free through rapid pulse truncation, and retained thanks to the impeding effect of the helium environment on molecular rotation~\cite{toennies_superfluid_2004}. We illustrate the opportunities that field-free aligned molecules open by measuring the alignment-dependent strong-field ionization yield of a thiophene oligomer. Our technique will enable molecular-frame experiments, including ultrafast excited state dynamics, on a variety of large molecules and complexes.}

\begin{figure}
    \centering
   \includegraphics[width = 88 mm]{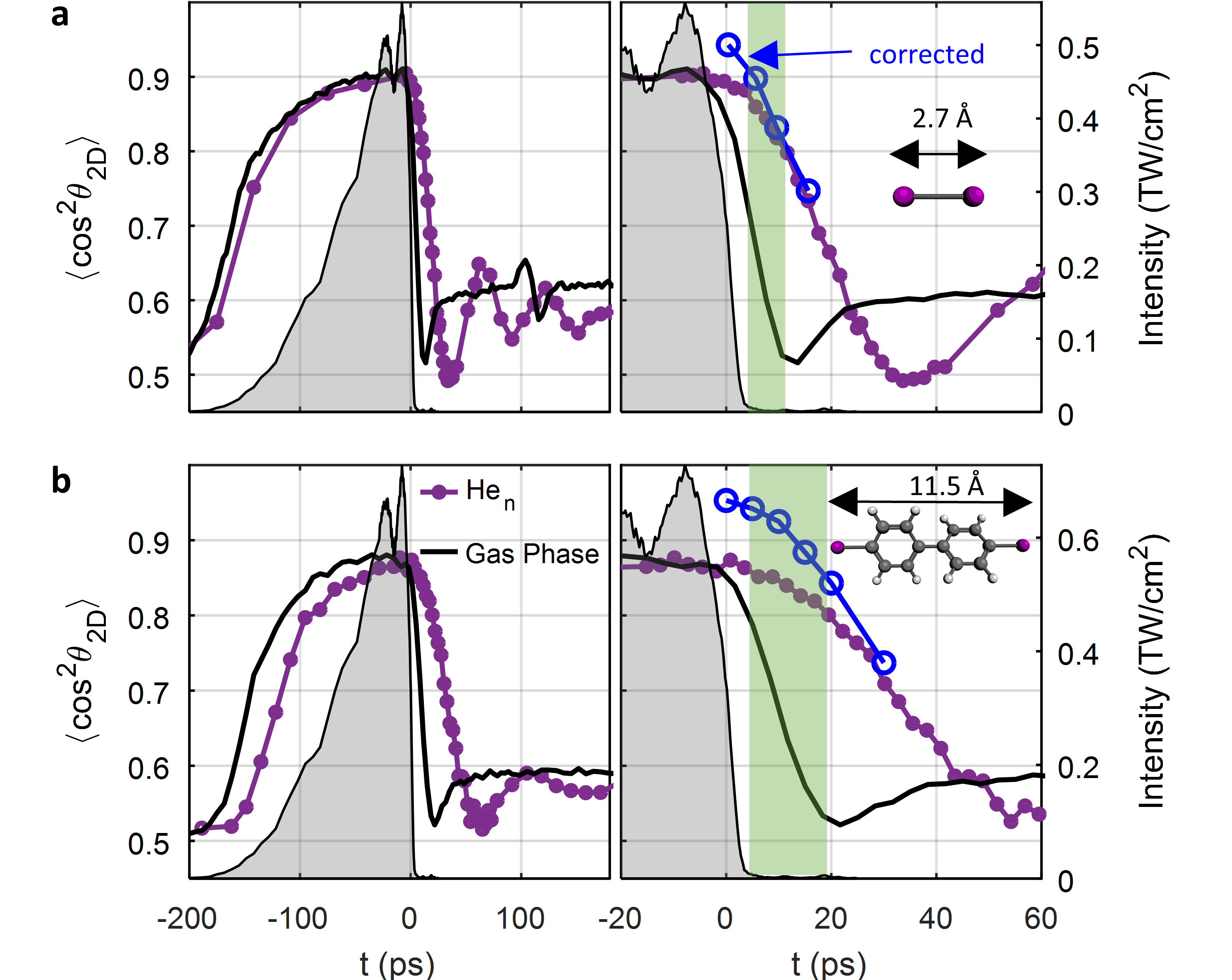}
 \caption{\textbf{1-dimensional alignment of \ce{I_2} and diiodobiphenyl molecules.}
Alignment dynamics of \ce{I_2} (panels \textbf{a}) and 4,4’-diiodobiphenyl (panels \textbf{b}) molecules, represented by $\cost$ and recorded as a function of time. The measurements are conducted both for isolated molecules (black curves) and for molecules in He droplets (purple curves) under identical laser conditions. The intensity profile of the truncated alignment pulse is shown by the grey shaded area and refers to the right vertical axis. The panels on the right show a zoom of the post-truncation region. The green shaded areas in these panels mark the intervals where the alignment field intensity is $<$~1~\% of its peak and $\cost \geq 0.80$. The structure of each molecule is inset, along with a scalebar representing the I--I bond length. In droplets, the values for \cost after correction for non-axial recoil at selected times are also shown as blue open circles.}

  \label{fig1}
\end{figure}

The first step of our technique is to induce strong molecular alignment with a laser pulse turned on (10\% -- 90\%) in 100 ps. The temporal evolution of the alignment is measured through timed Coulomb explosion, triggered by an intense 40 fs probe pulse sent at time $t$, and recording the emission direction of fragment ions (see Methods). The purple curve in \autoref{fig1}a depicts the result for 1-dimensional alignment of iodine (\ce{I_2}) molecules inside He droplets using a linearly polarized alignment pulse. The degree of alignment, $\cost$, rises from 0.5, the value characterizing random alignment, before the pulse to 0.90 at the peak of the pulse, with $\theta_\text{2D}$ denoting the angle between the projection of the recoil direction of an \ce{I^+} fragment ion and the polarization of the alignment pulse~\cite{shepperson_strongly_2017}.  For \ce{I_2} molecules perfectly aligned along the alignment pulse polarization $\cost$ would equal 1. The gradual increase of $\cost$ along with the alignment pulse intensity shows that the alignment process essentially evolves adiabatically in accordance with previous experiments~\cite{shepperson_strongly_2017}.

The second step is rapid truncation of the laser pulse at its peak to create field-free alignment, implemented by spectral truncation of a chirped pulse with a longpass optical filter~\cite{Adam_truncated_2018}. The alignment laser intensity drops by more than a factor of 100 over $\sim$~10 ps, however the degree of alignment decreases at a much slower rate. The right panel of \autoref{fig1}a shows that $\cost$ retains a value between 0.86 and 0.80 from $t$ = 5 ps to $t$ = 11 ps. In this 6 ps-long time window, marked by the shaded green area, the alignment pulse intensity is less than 1~\% of the value at the peak and the molecules are still well-aligned.
By comparison, isolated molecules in a cold molecular beam attain a similar value of $\cost$ at the peak of the laser pulse, but after truncation $\cost$ drops much faster and in the 5-11 ps interval $\cost$ is reduced to 0.69-0.52 (black curve in \autoref{fig1}a). The rapid decrease of $\cost$ in isolated \ce{I_2} is due to rotational dephasing characterizing freely rotating molecules~\cite{underwood_switched_2003, Adam_truncated_2018}. By contrast, in the droplets the He environment impedes the free rotation of the molecules~\cite{shepperson_strongly_2017,pentlehner_laser-induced_2013}, which comes to our advantage in terms of granting a period of field-free alignment after the pulse is switched-off.

An additional advantage of molecules in He droplets is that they have a rotational temperature of only 0.4 K~\cite{toennies_superfluid_2004}. In general, this leads to stronger alignment compared to gas-phase molecules~\cite{shepperson_strongly_2017} because the latter typically cannot be cooled to such low temperatures, in particular not the larger molecules mentioned below. In fact, the true degree of alignment is better than the $\cost$ values stated so far because the initial recoil direction of the \ce{I^+} ions, defined by the alignment of their parent molecule, is blurred due to collisions with He atoms on the way out of the droplet towards the detector. This effect can be corrected for~\cite{christensen_deconvoluting_2016} and we find that $\cost$ is actually between 0.90 and 0.84, shown as blue circles in \autoref{fig1}, in the 5-11 ps field-free interval.

Next, we turn to a much larger molecule, 4,4’-diiodobiphenyl (DIBP). The linearly polarized laser pulse is expected to align the most polarizable molecular axis, which is the I-I axis, along its polarization, i.e. a situation similar to that of the \ce{I_2} molecules. Again, $\cost$ is obtained from recording the emission direction of the Coulomb exploding \ce{I^+} ions. The results, displayed in \autoref{fig1}b, show that for DIBP in He droplets $\cost$ reaches 0.87 (0.95) shortly before truncation and 0.86-0.80 (0.94-0.84) between t = 5 and 20 ps, where the laser intensity is reduced to $<$~1~\% of its maximum value.  (The numbers in parentheses are the $\cost$ values obtained after correction for non-axial recoil effects and scattering on He atoms~\cite{christensen_deconvoluting_2016,shepperson_strongly_2017}.) For the isolated DIBP molecules such an interval of strong, field-free alignment is not present. Compared to \ce{I_2}, DIBP has a much higher moment of inertia, and the degree of alignment decays slower after pulse truncation. This points to the fortuitous effect that in general, the larger the molecule is, the longer it can be field-free aligned in the He droplets (compare the shaded green areas in \autoref{fig1}a and b).

\begin{figure}
    \centering
   \includegraphics[width = 88 mm]{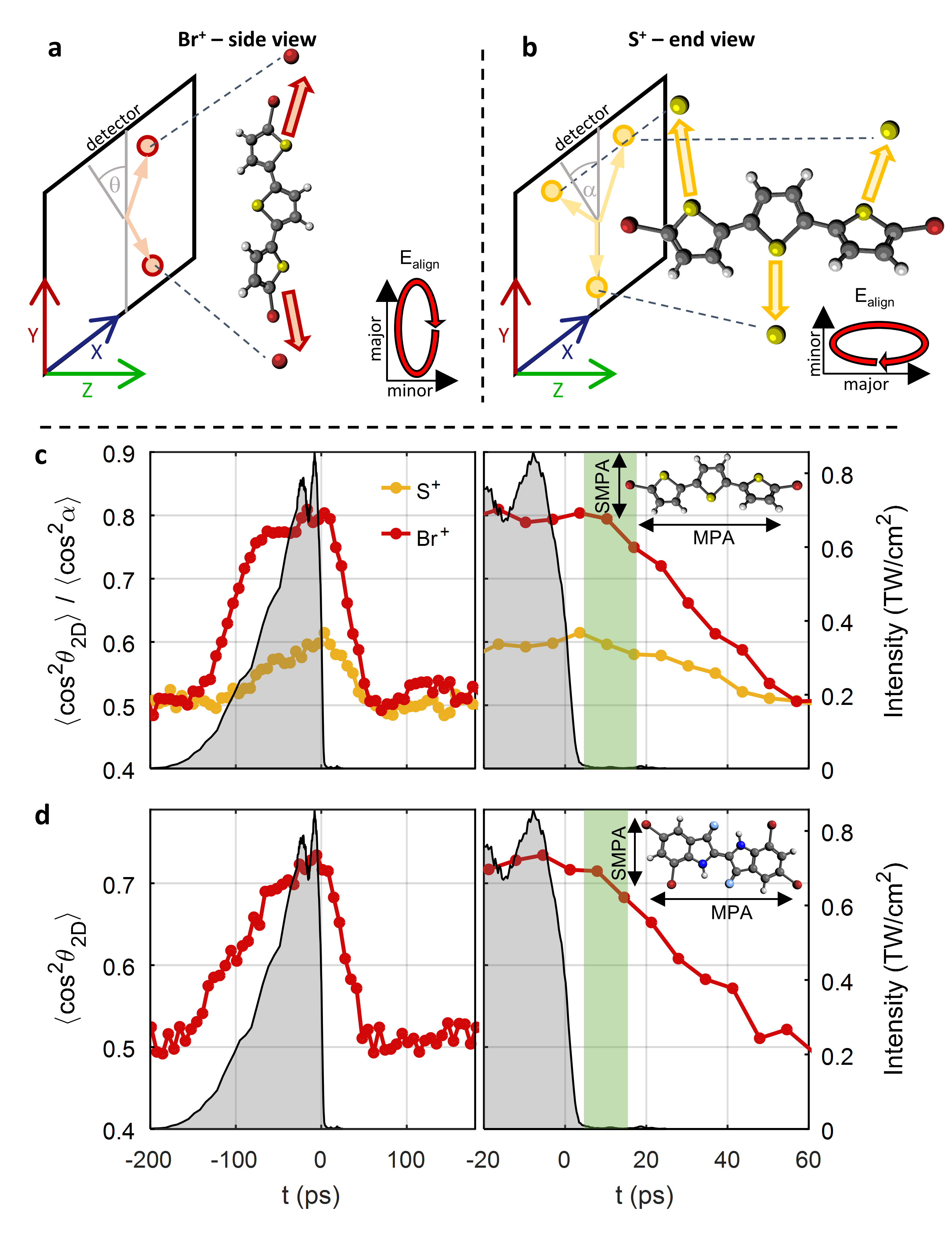}
 \caption{\textbf{3-dimensional alignment of dibromoterthiophene and tetrabromoindigo molecules in He droplets.}
\textbf{a}, \textbf{b}, Illustration of how 3D alignment of DBT is characterized. In \textbf{a} the major polarization axis of the alignment pulse is directed along the Y-axis. In this side-view of the molecules their most polarizable axis (MPA) is confined along the Y-axis and its alignment is characterized by $\cost$ where $\theta_\text{2D}$ is the angle between the emission direction of a \ce{Br^+} ion and the Y-axis.
In \textbf{b} the major polarization axis of the alignment pulse is directed along the Z-axis. In this end-view of the molecules their MPA is confined along the Z-axis and their second most polarizable axis (SMPA) along the Y-axis. The alignment of the SMPA is characterized by $\cosa$ where $\alpha$ is the angle between the emission direction of a \ce{S^+} ion and the Y-axis. \textbf{c}, The time dependence of $\cost$ and $\cosa$ for DBT molecules induced by an elliptically polarized alignment pulse with an intensity ratio of 3:1. \textbf{d}, The time dependence of $\cost$ for TBI molecules induced by an elliptically polarized alignment pulse with an intensity ratio of 3:1. These data were recorded in the side-view and as for DBT $\theta_\text{2D}$ refers to the angle between the MPA and the major polarization axis. For both \textbf{c} and \textbf{d} the right vertical axis gives the intensity of the alignment pulse. The panels on the right show a zoom of the post-truncation region and the green shaded areas mark the field-free aligned intervals where \cost has dropped $< 10\%$. The MPA and SMPA are overlaid on the molecular structures.}
  \label{fig2}
\end{figure}

To demonstrate the generality of our technique, notably towards complex systems, we performed experiments on two more molecules. The first, 5,5''-dibromo-2,2':5',2''-terthiophene (DBT), is an oligomer of three thiophene units and can be thought of as a prototype of polythiophenes used in molecular electronics~\cite{facchetti_conjugated_2011}. The alignment pulse is elliptically polarized with the purpose of inducing 3D alignment~\cite{larsen_three_2000}. Therefore, we expect that the most polarizable axis (MPA) aligns along the major polarization axis simultaneously with the second most polarizable axis (SMPA) aligning along the minor polarization axis, see \autoref{fig2}a,b. The alignment dynamics recorded for DBT, shown in \autoref{fig2}c, confirm these expectations. Around the peak of the pulse $\cost$ ($\cosa$) for the \ce{Br^+} (\ce{S^+}) ions reaches a maximum of $\sim$0.80 ($\sim$0.60) showing that the MPA and the SMPA are aligned simultaneously and, therefore, that the molecule is 3D aligned (see Supplementary Information for details).  This is consistent with recent demonstrations of 3D alignment for smaller molecules in He droplets~\cite{chatterley_three-dimensional_2017}. The $\cost$ and $\cosa$ values observed here are lower than in previous works because the fragment ions detected, notably the \ce{S^+} ions, do not recoil directly along the aligned axes as illustrated in \autoref{fig2}a,b. This is a natural consequence of the complex structure and low symmetry of DBT, and so, unlike the cases of \ce{I_2} and DIBP, $\cost$ can only provide a qualitative rather than a quantitative measure for the degree of alignment.

The crucial finding is, however, the lingering of the alignment after pulse truncation. From $t$ = 5 to 24 ps, where the laser intensity is $<$ 1\% of the peak value, $\cost$ for the \ce{Br^+} ions drops only from 0.80 to 0.72, and \cosa remains $>0.58$ for the \ce{S^+} ions. This demonstrates a 19 ps-long interval, marked by the green area, where the molecules are 3D aligned under field-free conditions. The second complex molecule studied was 5,7,5'7'-tetrabromoindigo (TBI), a brominated derivative of the indigo dye responsible for the colour in blue jeans. \Autoref{fig2}b shows that it is also possible to 3D align this molecule for about \SI{17}{ps} under field-free conditions. Details are given in the Supplementary Material.

To assess the field-free nature of the alignment created we measured the yield of intact parent cations, here created by strong-field ionization with the probe pulse. If the parent ion is created in the presence of an alignment pulse, which is the case in the adiabatic alignment regime, the parent ion will subsequently be exposed to the strong alignment field during its turn-off, which typically lasts \SI{>100}{ps}. While the alignment pulse is non-resonant for the neutral parent molecules, the cations of most large molecules will absorb at the alignment laser wavelength (800 nm). As a consequence, the parent ion is likely to absorb one or several photons from the alignment pulse, leading to dissociation or even further ionization. This destruction of the parent ion has been observed in several previous experiments~\cite{boll_imaging_2014,slater_coulomb-explosion_2015} and precluded the use of the parent ions as experimental observables.

\begin{figure}
    \centering
   \includegraphics[width = 88 mm]{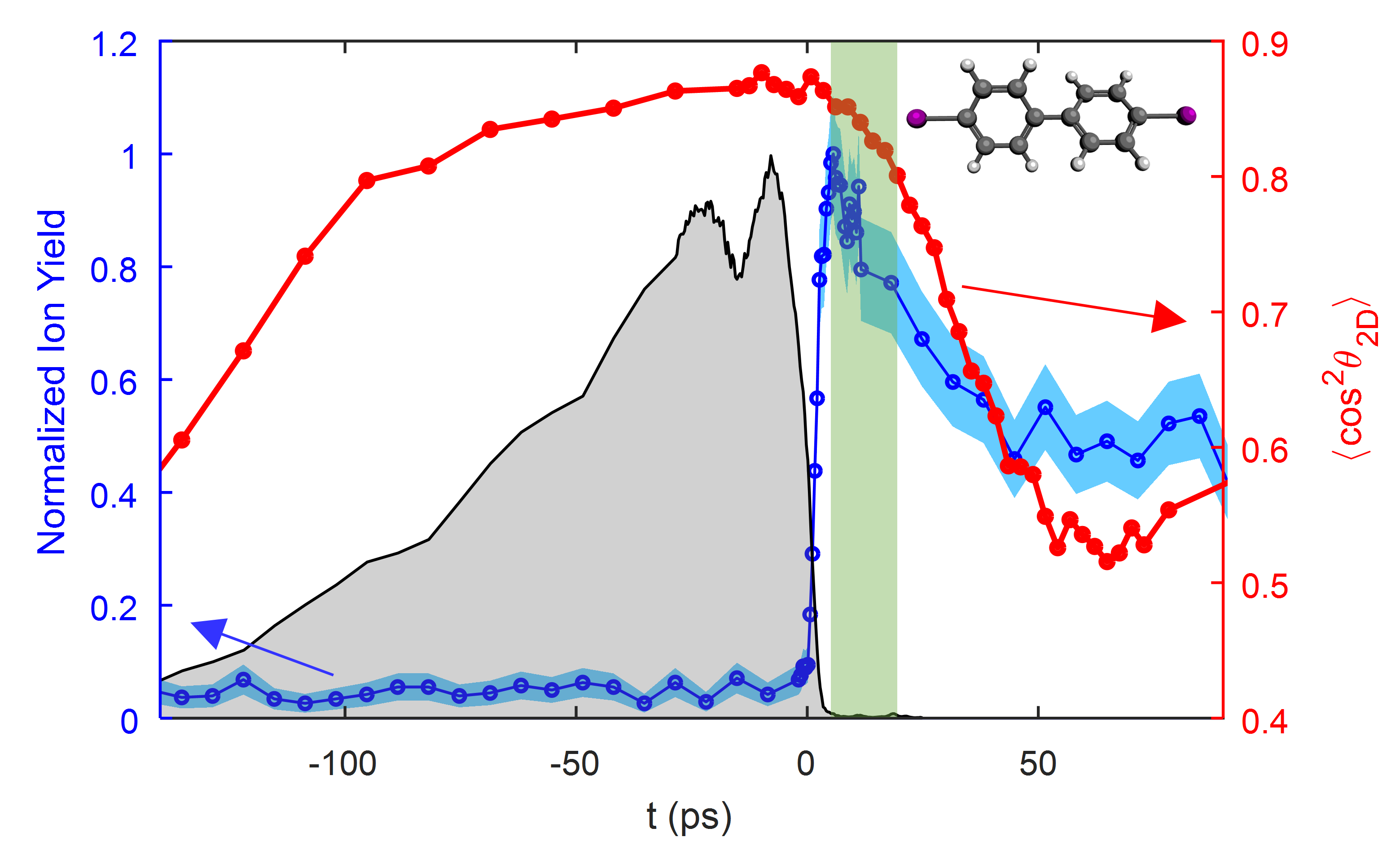}
 \caption{\textbf{Time-dependent yield of intact parent ions from strong-field ionization of diiodobiphenyl molecules in He droplets.}
 The DIBP molecules are 1D aligned. The 800 nm ionization pulse is 40 fs long (FWHM), has an intensity of \SI{2.4e14}{W/cm^2} and is linearly polarized perpendicular to the polarization axis of the alignment pulse. The blue shaded area represents the 95\% confidence intervals for the parent signal. The time-dependent degree of alignment is shown by the red curve.}
  \label{fig3}
\end{figure}

\Autoref{fig3} shows the yield of parent ions created by ionization of 1D aligned DIBP molecules with a strong-field probe pulse sent at time $t$. When the probe pulse arrives during the alignment pulse the parent ion signal is almost zero but when it arrives after the pulse the parent ion signal increases sharply and reaches a maximum at $t \sim 7$ ps. The sudden increase of the parent signal by almost a factor of 20 shows that the truncation reduces the alignment pulse intensity sufficiently to prevent destruction  of the parent ions. Importantly, the maximum of the parent ion yield occurs at a time where the molecules are still strongly aligned as can be seen from the $\cost$ curve, reproduced on the figure. As such \autoref{fig3} demonstrates that our method enables ionization experiments on sharply aligned molecules, without the alignment field destroying the fragile parent ions.  Note that after the maximum the parent ion yield decreases along with the degree of alignment because the strong-field processes involved depend on how the molecules are turned in space~\cite{pavicic_direct_2007,hansen_orientation-dependent_2012}.

\begin{figure}[ht]
    \centering
 \includegraphics[width = 88 mm]{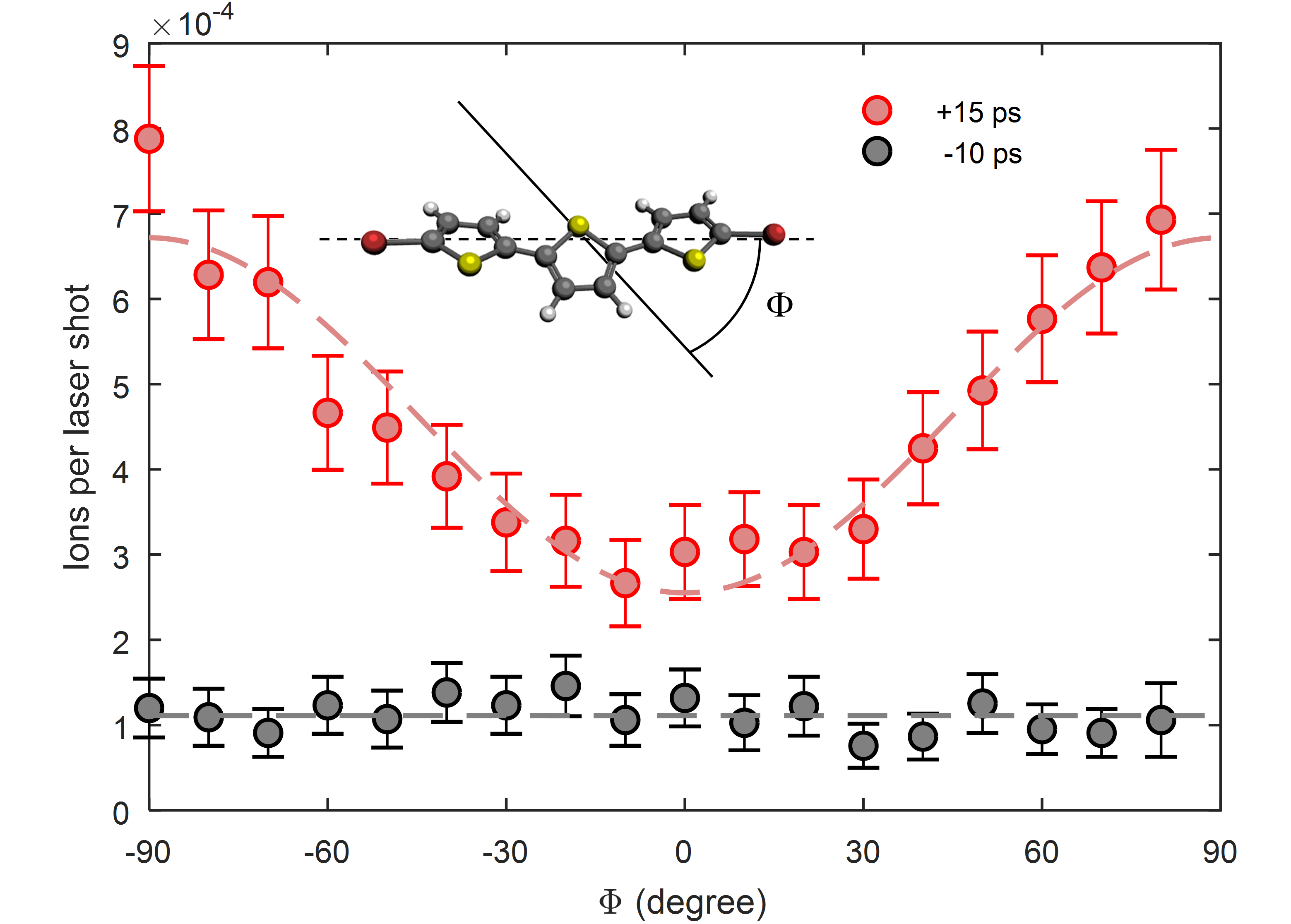}
 \caption{\textbf{Alignment-dependent strong-field ionization yields of dibromoterthiophene molecules in He droplets.}
DBT molecules are 1D aligned by a linearly polarized alignment pulse and ionized by a linearly polarized probe pulse (same parameters as mentioned in the caption to \autoref{fig3}). The yield of intact \ce{DBT^+} parent ions is measured as a function of the angle, $\Phi$, between the polarization directions of the two laser pulses - illustrated on the sketch of the molecular structure. The measurements are performed at two times, around the peak of the alignment pulse ($t= \SI{-10}{ps}$) and in the field-free window ($t= \SI{15}{ps}$). The dashed lines serve to guide the eye.}
  \label{fig4}
\end{figure}

Finally, we performed a simple proof-of-principle experiment to demonstrate the possibilities opened by the field-free alignment period for large molecules. The example chosen was a prototype linear dichroism experiment~\cite{dong_vibrational_2002}, the alignment-dependent ionization yield of DBT molecules, a phenomenon that has been subject to many studies for smaller molecules in gas-phase~\cite{pavicic_direct_2007,hansen_orientation-dependent_2012,luo_multiorbital_2017}. The DBT molecules were 3D aligned and strong-field ionized by a linearly polarized probe pulse sent at $t$ = 15 ps. The yield of the \ce{DBT^+} ions was recorded as a function of the angle between the probe pulse polarization and the major polarization axis of the alignment pulse, which held the MPA, approximately coinciding with the Br-Br axis (see \autoref{fig2}), fixed-in-space. The results, represented by the red circles in \autoref{fig4}, show that the detection of intact \ce{DBT^+} ions more than doubles when the probe pulse is polarized perpendicular instead of parallel to the MPA axis. A detailed analysis of the alignment dependence is beyond the scope of this work. Rather, the key point is that our measurement would have been impossible in the presence of the alignment field as illustrated by the black-circle data points recorded at the peak of the alignment pulse ($t$ = - 10 ps). Here the alignment dependence of the ionization yield is not visible at all.

We have shown that a slowly turned-on, rapidly turned-off laser pulse provides a general approach to sharply align large molecules inside He nanodroplets under conditions that are effectively field-free for a period of at least 10-20 ps. We demonstrated that the residual alignment field in this period is so weak that it leaves fragile molecular parent ions unaffected. The same will be the case for molecules in electronically excited states, which opens unexplored possibilities for fs time-resolved imaging of molecules undergoing fundamental photo-induced intramolecular processes in electronically excited states using Coulomb explosion, linear dichroism, or diffraction by ultrashort x-ray or electron pulses~\cite{gomez_shapes_2014,tanyag_communication:_2015,zhang_communication:_2016,yang_diffractive_2016}. The long-lasting field-free period ensures that there is enough time to watch how unimolecular processes, like isomerization, evolve through their entire duration.
Moreover, the unique opportunities for building molecular complexes inside He droplets~\cite{choi_infrared_2006,yang_helium_2012} provides great potential for applying molecular frame measurements to image such systems and their dynamics. In particular, recording of intact parent ions following ionization of sharply aligned molecules by (intense) laser pulses will extend recent femtosecond laser-based methods to determine the structure of molecular dimers in He droplets from small linear systems to large complex systems for instance polycyclic aromatic hydrocarbon  like tetracene and pentacene. This should allow real-time imaging of intermolecular processes such as exciplex formation~\cite{miyazaki_real_2015}, singlet fission~\cite{smith_singlet_2010} and bimolecular reactions~\cite{cheng_femtosecond_1996}.



\section*{Methods}
The experiment used a helium droplet apparatus~\cite{shepperson_strongly_2017} and truncated alignment pulses as described previously~\cite{Adam_truncated_2018}. Helium nanodroplets were produced by continuously expanding 25 bar helium into vacuum through a \SI{5}{\micro m} nozzle, cooled to 14 K (\ce{I_2}), 13 K (DIBP \& DBT) or 12 K (TBI). Molecules were introduced into the droplets by sending them through a pickup cell filled with molecular vapour, introduced via either a leak valve (\ce{I_2}) or an in-vacuum oven (DIBP, DBT \& TBI). In all cases, the vapour pressure in the pickup cell was adjusted to optimize single molecule doping. The doped helium droplets entered the target region, where they were perpendicularly intersected by the spectrally truncated, chirped alignment pulse ($\lambda_{centre} = \SI{800}{nm}$, $\omega_0 = \SI{38}{\micro m}$, peak intensity $I_{align} \sim \SIrange{6e11}{9e11}{W / cm^2}$) and a time delayed probe pulse ($\lambda_{centre} = \SI{800}{nm}$, duration \SI{40}{fs} FWHM, $\omega_0 = \SI{25}{\micro m}$, $I_{probe} \sim \SI{2.4e14}{W/cm^2}$). The probe spot size was considerably smaller than the alignment to minimize focal volume effects. Unless otherwise stated, the alignment laser pulse was polarized with the major axis parallel to the detector, and the probe beam was linearly polarized perpendicular to it.

Degrees of alignment were characterized by Coulomb explosion imaging, and detecting the velocity vectors of charged fragments with a velocity map imaging spectrometer, gated in time such that it was sensitive only to a single ion of interest. Only the ions with higher kinetic energies were considered for \cost determinations. Parent ion yields were also detected using the VMI spectrometer, except only ions with very low kinetic energy were considered. In both cases, there is an unavoidable background from hot effusive molecules of a few percent, however this does not affect the conclusions. The contribution of these effusive molecules is removed in the non-axial recoil correction shown in Fig. 1.

Gas phase \ce{I_2} and DIBP molecules were investigated using the same experimental setup, except that the molecules were sourced from an Even-Lavie pulsed valve, backed by \SI{80}{bar} of He seeded with \ce{I_2} or DIBP.

\section*{Acknowledgements}
We acknowledge support from the European Research Council-AdG (Project No. 320459, DropletControl). Also, this research was undertaken as part of the ASPIRE Innovative Training Network, which has received funding from the European Union's Horizon 2020 research and innovation programme under the Marie Sklodowska-Curie grant agreement No 674960.

\section*{Author contributions}
A.S.C, C.S, L.C., B.S, M.H.R. designed and performed experiments.
A.S.C, C.S, L.C., H.S. analyzed the experimental results.
All authors took part in regular discussions and were involved in the completion of the manuscript.

\section*{Supplementary Material}
Provides additional details and information about 3D alignment of dibromoterthiophene and tetrabromoindigo molecules.

\bibliographystyle{naturemag}

\bibliography{Zotero-June05}

\clearpage

\clearpage

\end{document}